\documentclass[prb,aps,preprint,superscriptaddress]{revtex4}

\usepackage[dvips]{epsfig}
\usepackage{float}
\usepackage{amsmath}
\usepackage{amssymb}
\usepackage{amsfonts}
\usepackage{enumerate}
\usepackage{color}
\usepackage[sort&compress]{natbib}
\usepackage{hyperref}

\begin{document}

\title{Hybrid Lithium Niobate and Silicon Photonic Waveguides}
\author{Peter O. Weigel}
\email{pweigel@eng.ucsd.edu}
\author{Marc Savanier}
\affiliation{Department of Electrical \& Computer Engineering, University of California, San Diego,\\ La Jolla, California 92093, USA}
\author{Christopher DeRose}
\author{Andrew T. Pomerene}
\author{Andrew L. Starbuck}
\author{Anthony L. Lentine}
\affiliation{Sandia National Laboratory, Applied Microphotonic Systems, Albuquerque, New Mexico 87185 USA}
\author{Vincent Stenger}
\affiliation{SRICO Inc., 2724 Sawbury Blvd Columbus, OH 43235-4579 USA}
\author{Shayan Mookherjea}
\email{smookherjea@ucsd.edu}
\affiliation{Department of Electrical \& Computer Engineering, University of California, San Diego,\\ La Jolla, California 92093, USA}

\date\today
\maketitle

\section{Introduction}
Lithium Niobate (LN) was once considered the most promising of materials for integrated optics\cite{0034-4885-56-3-001}, but despite a rich set of properties, the technology of LN integrated optics has not evolved as much as integrated optics in III-V semiconductors and silicon (Si) photonics\cite{smit2012moore,hochberg2010towards}. Although high performance stand-alone LN devices\cite{sohler2008integrated,Chiba2010} and a technique of ion-sliced thin-film LN\cite{levy1998fabrication,rabiei2004optical,Poberaj2010} have been developed over the past two decades, the technology of LN integrated optics has continued to rely on traditional waveguide fabrication techniques based on ion-exchange\cite{tamir1988guided}, diffusion and serial writing\cite{PSSA201026452} or mechanical sawing\cite{Ulliac20112417,Takigawa:14}, all of which are very different from the modern lithographic techniques and foundry processing available in Si or III-V photonics. 

We have developed a photonic waveguide technology based on a bi-material core, in which light is controllably and repeatedly transferred back and forth between sub-micron thickness crystalline layers of Si and LN bonded to one another, where the former is patterned and the latter is not\cite{patent-SM-MS}. The benefit of this approach is that deep ultraviolet (DUV) lithographic fabrication of Si photonic features is more precise, scalable and manufacturable than the traditional waveguide-fabrication techniques in LN. Since LN is not a CMOS-compatible material and is not accepted in CMOS foundries, we developed designs which allow all the waveguiding circuitry to be defined in Si, requiring only a single back-end process step of bonding LN to form the functional chip. Using two different guided modes and an adiabatic mode transition between them, we demonstrate a set of building blocks such as waveguides, bends, and couplers which can be used to route light underneath an unpatterned slab of LN, thus enabling complex and compact lightwave circuits in LN with fabrication ease and low cost.  

Our approach demonstrated here to forming waveguides in LN is based on a two-material cross-section, consisting of few-hundred nanometer single-crystal thin films of Si and LN bonded face-to-face. Si and LN are two important materials for integrated optics which have significantly different coefficients of thermal expansion. The technology of bonding Si to LN has been studied in the past decade\cite{howlader2006room,Takigawa2006,Lee:11,Chen:14,Chiles:14}, with the key differences of our approach explained here. Bonding achieves much higher quality and better optical properties than growth of LN on Si using sol-gel processes\cite{Yoon96} or chemical vapor deposition\cite{Sakashita}. Both manufacturer-provided Si and LN wafers have a $\text{SiO}_\text{2}$ cladding layer which is a few microns thick, and a handle material (Si and LN, respectively) for the substrate which is several hundred microns thick. 

We first pattern all features required for waveguiding in the Si wafer alone, using deep ultraviolet (DUV) lithography. This allows much finer features to be patterned than is possible in LN (which has previously constrained LN waveguides to be multi-moded\cite{guarino2007electro}). In our approach, the LN wafer is conveniently left blank and unpatterned, in contrast to etching LN\cite{ConfHuiECOC2008,Rabiei-etching}, a technique that is less mature than etching silicon, which is available in every foundry process. Upon bonding, a hybrid waveguide is formed, in which the optical mode is distributed between the Si and LN layers in a controllable ratio, while remaining in the transverse-electric (TE) polarized single-mode regime, which is highly desirable for Si photonics. As will be discussed below, light propagates mainly in the silicon layer in certain sections of the layout, and primarily in the LN layer where desired and makes transitions between the two layers at several locations. We can also design devices that are outside the LN-Si bonded region and thus behave entirely like conventional Si photonics. In this way, the technology for Si photonics can be leveraged to enhance LN integrated optics, and vice versa. 

The main advantages of this approach are: (a) the requirements on LN fabrication are reduced to a minimum, i.e.,~a single bonding step to the manufacturer-supplied wafer after all silicon patterning has completed; and (b) both the cross-sectional mode area and the length of components are greatly reduced compared to traditional LN ion-exchanged or diffused waveguides, allowing complex circuits to be realized in a small area. By not patterning, etching or sawing LN, its material properties are kept pristine\cite{han2015optical}. The tradeoffs are that: (a) the bonding step should be performed at a low temperature since Si and LN have different coefficients of thermal expansion, and (b) the resulting hybrid modes have a higher refractive index dispersion compared to low-index contrast (e.g., doped-glass) integrated optics, and thus, components will be optimized for a desired wavelength range, e.g., the telecommunications C-band (1530~nm--1565~nm).  

Fig.~\ref{F1} shows images of a chip whose dimensions are that of a typical field size of a stepper projection DUV lithography system (few centimeters squared). The Si photonic chip was fabricated by a foundry process in 150~nm thick Si (100)-oriented device layer, with  $\text{SiO}_2$ buffer of thickness $3\ \mathrm{\mu m}$ separating the device layer from the thick Si handle. The LN chip consists of an unpatterned x-cut LN film with nominal thickness of 750 nm supported by a $\text{SiO}_2$ buffer layer of thickness $2\ \mathrm{\mu m}$ and an LN handle of thickness 0.5~mm. We directly bonded the LN chip to the Si chip i.e., without any intervening polymeric `glue' layer\cite{guarino2007electro,Chen:12}, creating a hybrid material stack whose combined thickness (approximately $0.9\ \mathrm{\mu m}$) was sufficient to support the desired modes.  

\section{Fabrication} 
The silicon features were patterned using DUV lithography on 150~mm diameter silicon-on-insulator wafers with a substrate oxide thickness of $3\ \mathrm{\mu m}$ at Sandia National Laboratories. A blanket etch was used for height reduction of the device layer from 250~nm to 150~nm followed by patterning to define waveguides and bonding pads simultaneously (i.e., at the same height). The thin-film LN on insulator wafers were procured from a commercial source (NanoLN Jinan Jingzheng Electronics Co. Ltd.), with a nominal top LN thickness of 750 nm, silicon oxide buffer of thickness $2\ \mathrm{\mu m}$ and LN handle  of thickness 0.5~mm, diced into chips of size 2.1~cm $\times$ 1.7~cm. The material properties of such thin-films have been documented elsewhere\cite{han2015optical}.

\subsection{Bonding}
There are several techniques that could be used for bonding Si to LN, most of which rely on managing the temperature budget of the process, in view of the different thermal coefficients of expansion of the two crystalline materials. We used a room-temperature bonding process without any intermediate layers i.e., direct bonding. The cleaning process included a quick buffered oxide etch for the Si chip only (4 seconds), a piranha acid etch to remove organic contaminants (2:1 $\mathrm{H}_2\mathrm{SO}_4:\mathrm{H}_2\mathrm{O}_2$, 5 minutes), $\mathrm{O}_2$ plasma surface activation (45 seconds, Technics PEIIB, 200~W RF power, 200 mTorr pressure), and several cleaning steps using acetone, methanol and deionized water rinses. After drying, the chips were aligned manually such that the z-axis of the (x-cut) LN was perpendicular to the long axis (y) of the optical waveguides. Bonding was initiated at room temperature by gently pressing and sliding outwards from the center using plastic tweezers or the backside of a clean room swab. More advanced methods may be used in commercial development. After bonding, approximately 825~nm of $\text{SiO}_2$ were sputtered at room temperature (Denton 635, 400 W, 5 mT; growth rate approximately 275 nanometers of $\text{SiO}_2$ per hour).

\subsection{Bond-Strength Testing}
We performed a bond strength analysis using an Imada force gauge and test stand to measure tensile ``pull-apart'' force\cite{vallin2005adhesion}. The average bond strength measured over a few test chips was $0.2 \pm 0.09$~MPa. These results eliminate the possibility of the sputtered oxide being an ad-hoc ``glue'' holding the wafers together. Because we did not perform any elevated temperature anneal in this process, typical bond strengths were less than those achieved after annealing, but were sufficient for handling and testing the chips. Chips have remained bonded for more than a month without any special storage techniques. For commercial and proprietary applications, modifications of this technique may be developed with greater bond strength. 

\section{Waveguide Modes}

\begin{figure}[t]
	\centerline{\includegraphics[width=0.8\textwidth]{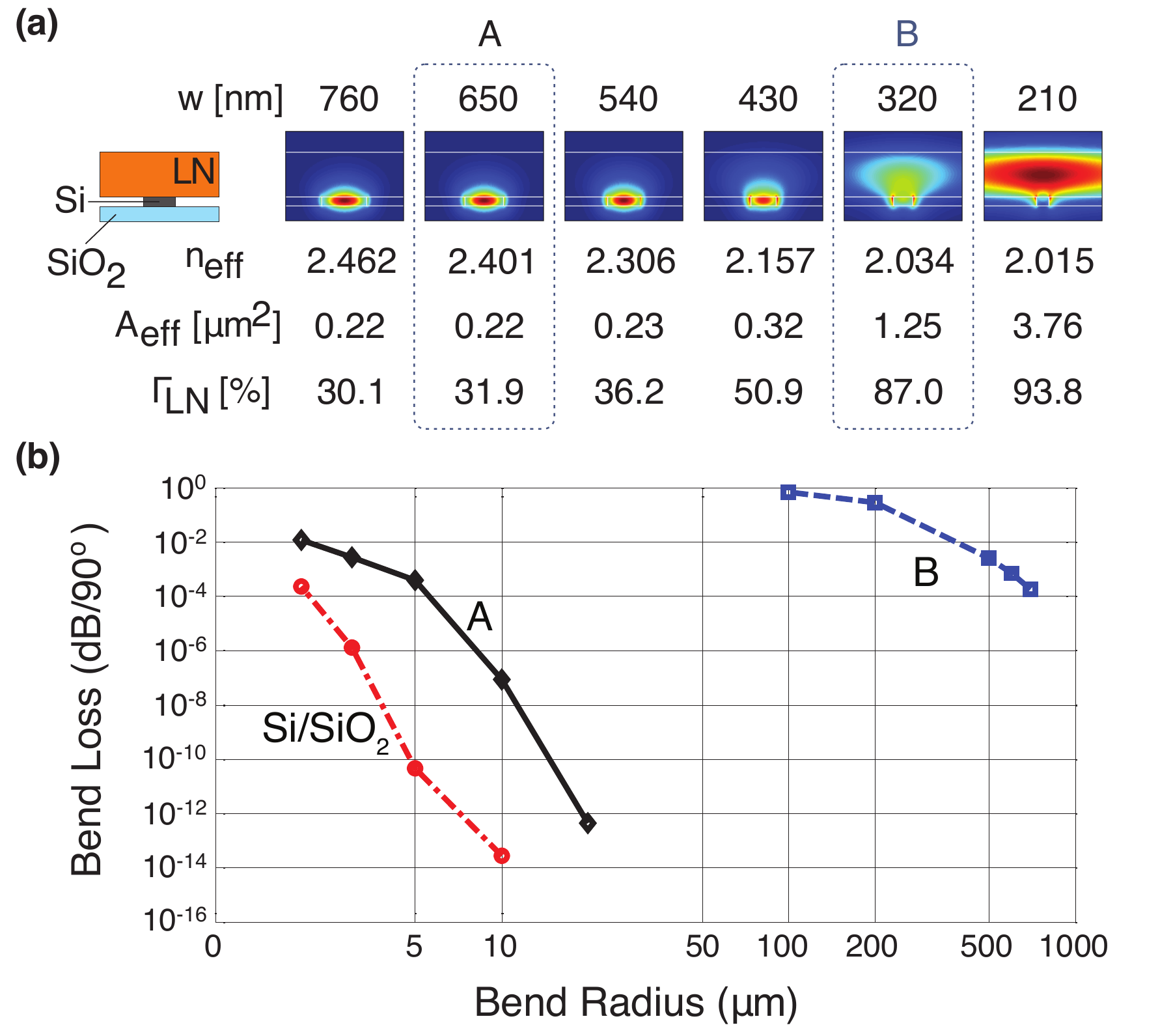}}
	\caption{(a) Calculated hybrid optical mode profiles  for different Si rib widths. The panels show the magnitude of the electric field in the TE polarization, with the E-field vector oriented along the crystal axis.  As the Si rib width $w$ decreases, the modal effective index ($n_\text{eff}$) decreases, the effective area ($A_\text{eff}$) increases, and the fraction of light in LN ($\Gamma_\text{LN}$) also increases. The dotted boxes indiciate the two cross-sectional designs (`A' and `B') used in the chip. (b) The calculated bending loss of the `A' cross-sectional mode is much lower than that of the `B' mode, and is not too different from the waveguides used in silicon photonics. Therefore, `A' is used for bends and compact routing, and `B' is used in straight waveguide sections when most of the light should ``see'' LN.}
	\label{F2}
\end{figure}

The dimensions of the Si features determine what fraction of the optical mode resides in Si and in LN. Two distinct waveguide cross-sections, labeled `A' and `B', are selected, as shown in Fig.~\ref{F2}a, and used in the microchip layout. These cross-sections vary only in the width of the Si structures, which allows for convenient fabrication in a single step of lithography. 

\begin{figure}[t]
	\centerline{\includegraphics[width=\textwidth]{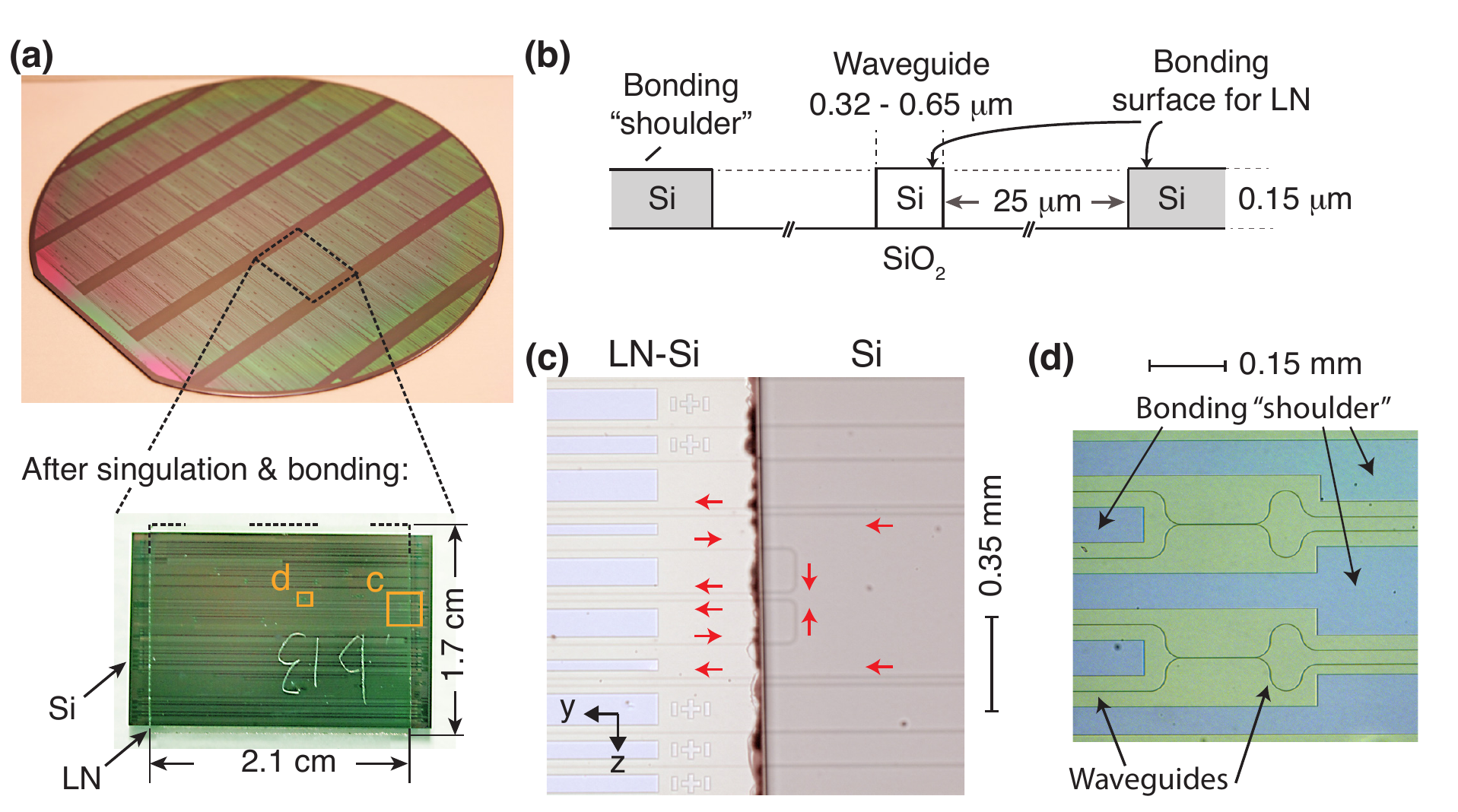}}
	\caption{(a) Silicon photonic components were fabricated using deep ultraviolet (DUV) lithography. Singulated dies (size: 25~mm $\times$ 16~mm) were bonded to diced pieces (size: 21~mm $\times$ 17~mm) of an unpatterned Lithium Niobate (LN)-on-insulator wafer. This particular LN chip was labeled `b13' by scratch marks on the LN substrate (0.5~mm thick, opposite to the bonded surface). (b) Schematic of the cross-section showing how waveguides and bonding ``shoulders'', which are at the same height as the waveguides, are conveniently formed in one lithographic etch step on the Si wafer. (c) Optical microscope image showing waveguides transitioning between the portion of the hybrid chip which is not covered by LN (i.e., conventional $\text{SiO}_2$-clad Si photonic waveguides), and that which is bonded to LN (and uses hybrid waveguides). Arrows colored red indicate the back-and-forth direction of light propagation in certain representative sections. (d) Mach-Zehnder interferometers defined in the LN-Si bonded section. }
	\label{F1}
\end{figure}

The numerical values of the widths and heights chosen here are suitable for wavelengths between 1530~nm and 1565~nm (i.e., the telecommunications C-band). Cross-section A consists of 650~nm wide and 150~nm tall Si features, bonded to LN. The lowest-order mode of the waveguide is quasi-TE-polarized and is similar to the quasi-$\text{TE}_0$ mode in conventional silicon photonics (see Fig.~\ref{F2}a). Wider waveguides are close to becoming multi-moded, which is undesirable. A waveguide using cross-section A can transition from an $\text{SiO}_2$-clad silicon photonic section to the bonded Si-LN region, as shown in Fig.~\ref{F1}c, with less than -0.3 dB loss calculated by numerical simulation (Lumerical software package). Furthermore, as shown by Fig.~\ref{F2}b, waveguides using cross-section A can support compact bends with radii of approximately $10\ \mathrm{\mu m}$ (similar to all-silicon photonics\cite{Vlasov:04}) whereas earlier demonstrations using ion-sliced LN waveguide bends have shown a radius of $100\ \mathrm{\mu m}$ \cite{guarino2007electro}. Some examples of compact bends underneath the LN chip are shown in Fig.~\ref{F1}c. 

Cross-section B consists of 320 nm wide Si features (same height as cross-section A) which are too small to support a waveguide mode by themselves, and require the bonded LN layer to support a well-defined mode at C-band wavelengths. Cross-section B is used when we want the light to interact with the crystal properties of LN; otherwise, cross-section A is used to route light on the hybrid chip. Therefore, it is desirable to maximize the fraction of light in LN for the mode in cross-section B. The confinement factor is defined as follows:  \newline $\Gamma_\text{LN} = \text{Re} \left[ \iint_{LN} \mathbf{E}\times\mathbf{H}^*\cdot \hat{y}\, dA\right]\ \bigg/ \ \text{Re} \left[\iint (\mathbf{E}\times\mathbf{H}^*\cdot \hat{y}, dA\right]$, where $\hat{y}$ is the direction of light propagation and the range of integration in the integral in the numerator is restricted to the LN region. In waveguides using cross-section A, $\Gamma_\text{LN} \approx 32\%$ whereas in waveguides using cross-section B, $\Gamma_\text{LN} \approx 90\%$. The transition between the two cross-sections is discussed below.  

A modified version of cross-section A, in which $\text{SiO}_2$ replaces LN as the upper cladding, is used outside the LN bonded region to demonstrate all-Si components made on the same platform and at the same device level. Another cross-section, with a reduced Si width of 180~nm is used for couplers at the edge of the Si chip, as commonly used in Si photonics\cite{Almeida:03}, and can also be fabricated in the same step. 

All these waveguide shapes are simple rectangles without relying on slots, tilted sidewalls or other difficult-to-fabricate shapes. There is no patterning in the LN layer which eliminates many of the challenges faced in the past\cite{guarino2007electro}. Conceptually, the waveguide structure is similar to that of the strip-loaded waveguide film studied in the 1970's\cite{Uchida:76}, but scaled to the deep sub-micron regime. The modal area of cross-section A (0.22 $\mathrm{\mu m}^2$) is, in fact, smaller, and that of cross-section B (1.25 $\mathrm{\mu m}^2$) is only slightly larger, than those of the smallest-area waveguides reported in LN, fabricated by the ion-milling technique\cite{Hu:09}.

As shown in Fig.~\ref{F1}(d), large rectangular areas were defined in the Si device layer at distances of about $30\ \mathrm{\mu m}$ from the waveguide edge. These ``bonding pads'' between the Si and LN chips are far from the waveguide core and serve no optical purpose. They are at the same height as the Si waveguides, since they are formed in the same lithographic step. They are similar to ``dummy fill'' features inserted to assist in chemical-mechanical polishing, but without a fragmented pattern.    

Although we do not fabricate active devices (e.g., modulators) in this work, simulations show that gold electrodes can be positioned directly on the LN thin-film layer (surface opposite to the bonded interface to Si) at a lateral distance of only $0.4\ \mathrm{\mu m}$ from the Si edge for cross-section A and $4\ \mathrm{\mu m}$ from the Si edge for cross-section B, for an estimated additional propagation loss of 0.1~dB/cm.

\begin{figure}[t]
	\centerline{\includegraphics[width=\textwidth]{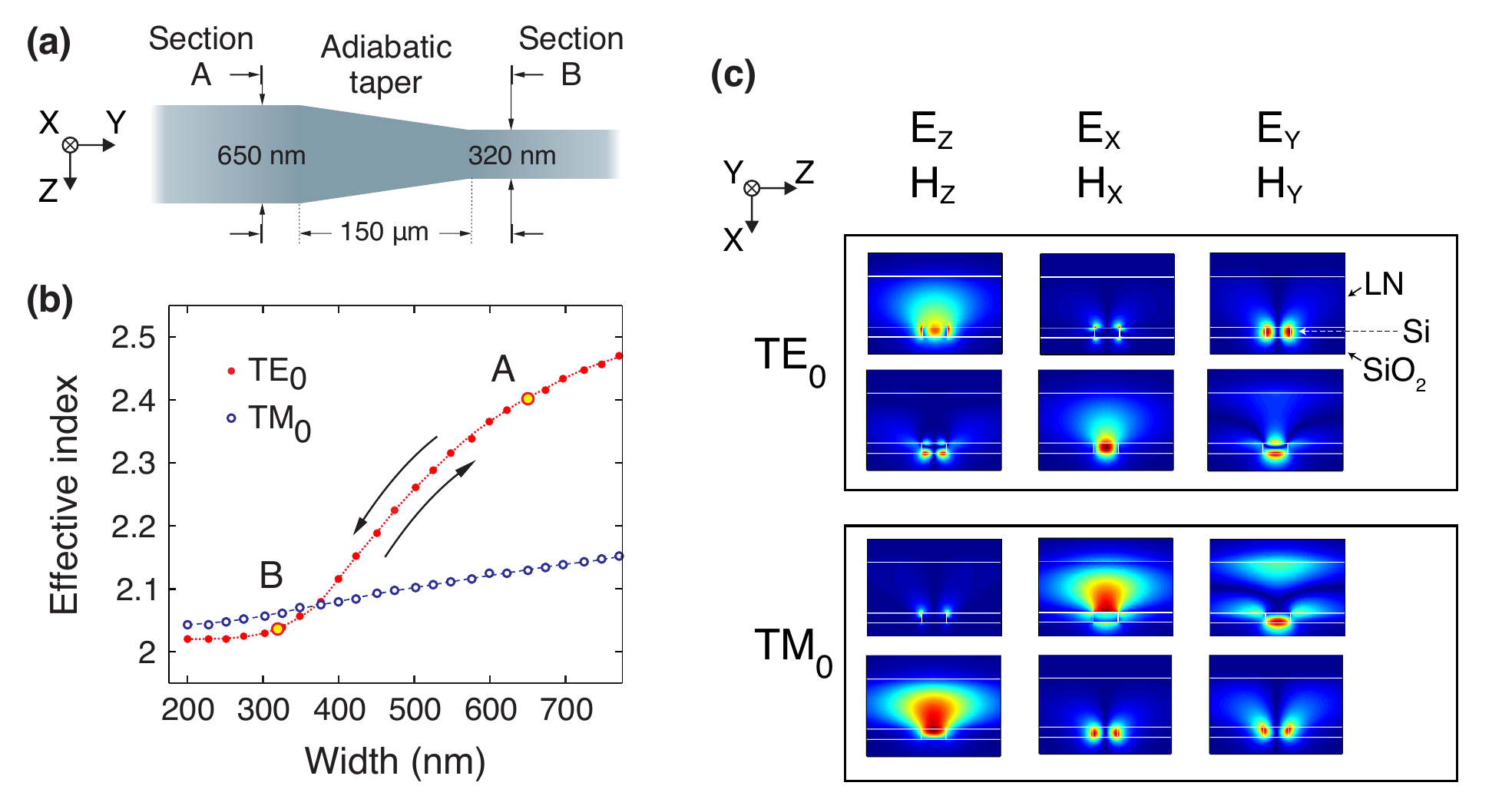}}
	\caption{(a) Gradual linear reduction of the Si rib width to transition between cross-sections A and B. (b) Numerical calculations of the modal effective index versus waveguide width, with yellow circles indicating the initial and final points of the taper. The TE-polarized mode does not hybridize with, or convert into, the TM-polarized mode in the taper, even though the latter becomes the fundamental mode at small widths. (c) The calculated magnitudes of the $E$- and $H$- field components at the crossing point, which show that field components involved in the dot-product in Eq.~(1) have opposite even-odd symmetry and therefore, are nearly orthogonal.}
	\label{F3}
\end{figure}

Along the direction of light propagation, adiabatic tapered transitions were defined in the Si layer (see Fig.~\ref{F3}a) as a linear change in the Si rib width over a distance of $150\ \mathrm{\mu m}$. Whereas the quasi-TE-polarized mode has the highest effective refractive index in cross-section A, it is actually the quasi-TM-polarized mode which has the higher refractive index in cross-section B. However, the light remains in the quasi-TE polarization as it traverses the taper because of the design of the waveguide taper used to convert back and forth between cross-sections A and B and serves to ``push'' and ``pull'' the optical mode to and from the LN layer. 

Leveraging symmetry principles, we are able to draw the waveguide down to a narrow width that pushes a very large fraction of the light into LN while maintaining its state of polarization all the way from the feeder waveguides. This is important because in conventional Si photonics (for example, as used outside of the LN bonded region), the quasi-TM-polarized modes are lossy and do not support compact bends, and thus the quasi-TE-polarized mode is used extensively\cite{Vlasov:04}.  

In our design, light is first guided using cross-section A, and the taper-induced coupling between the two fundamental (lowest-order) modes---the quasi-TE-polarized mode in cross-section A and the quasi-TM-polarized mode in cross-section B---is zero to first order. More specifically, the coupling coefficient between two modes $\mathbf{E}_1$ and $\mathbf{E}_2$ whose propagation constants are $\beta_1$ and $\beta_2$, respectively, is defined in the coupled-mode approximation by an integral over the waveguide cross-section (`c.s.')\cite{bk-Bures} 
\begin{equation}
C_{00}(y) \approx \frac{k}{4(\beta_1-\beta_2)} \sqrt{\frac{\epsilon_0}{\mu_0}} \iint_\text{c.s.} \mathbf{E}_2 \cdot \mathbf{E}_\text{1}^* \left[\frac{d}{dy}n(y)^2\right] \, dx\, dz
\end{equation}
where $k = 2\pi/\lambda$, $\lambda$ is the optical wavelength and $n(y)$ is the refractive index profile which varies along the direction of propagation ($y$). Since the refractive index is symmetric about $(z=0)$, the integral in Eq.~(1) vanishes for the modes shown in Fig.~\ref{F3}c because each term of the vectorial dot-product multiplies a function that has even symmetry about the center of the waveguide with one that has odd symmetry.  

In fact, compared to other hybrid Si-LN structures\cite{Rabiei-etching,Chen:14,rao2015heterogeneous}, our design achieves the highest TE-polarized mode-fraction in LN  while also providing a pathway for integration to Si photonics by first coupling light into cross-section A (which is similar to a traditional Si photonics cross-section), and then transitioning into cross-section B adiabatically. Attempting to couple from an external input to the quasi-TE mode in cross-section B would render the device very susceptible to roughness and bending losses, because it is not the fundamental mode. Widening the Si waveguide effectively pulls the light back into Si again, and, because it remains in the TE-polarization,  allows tight bends, compact directional couplers etc.   

\section{Building blocks of photonic circuits}
\begin{figure}[t]
	\centerline{\includegraphics[width=\textwidth]{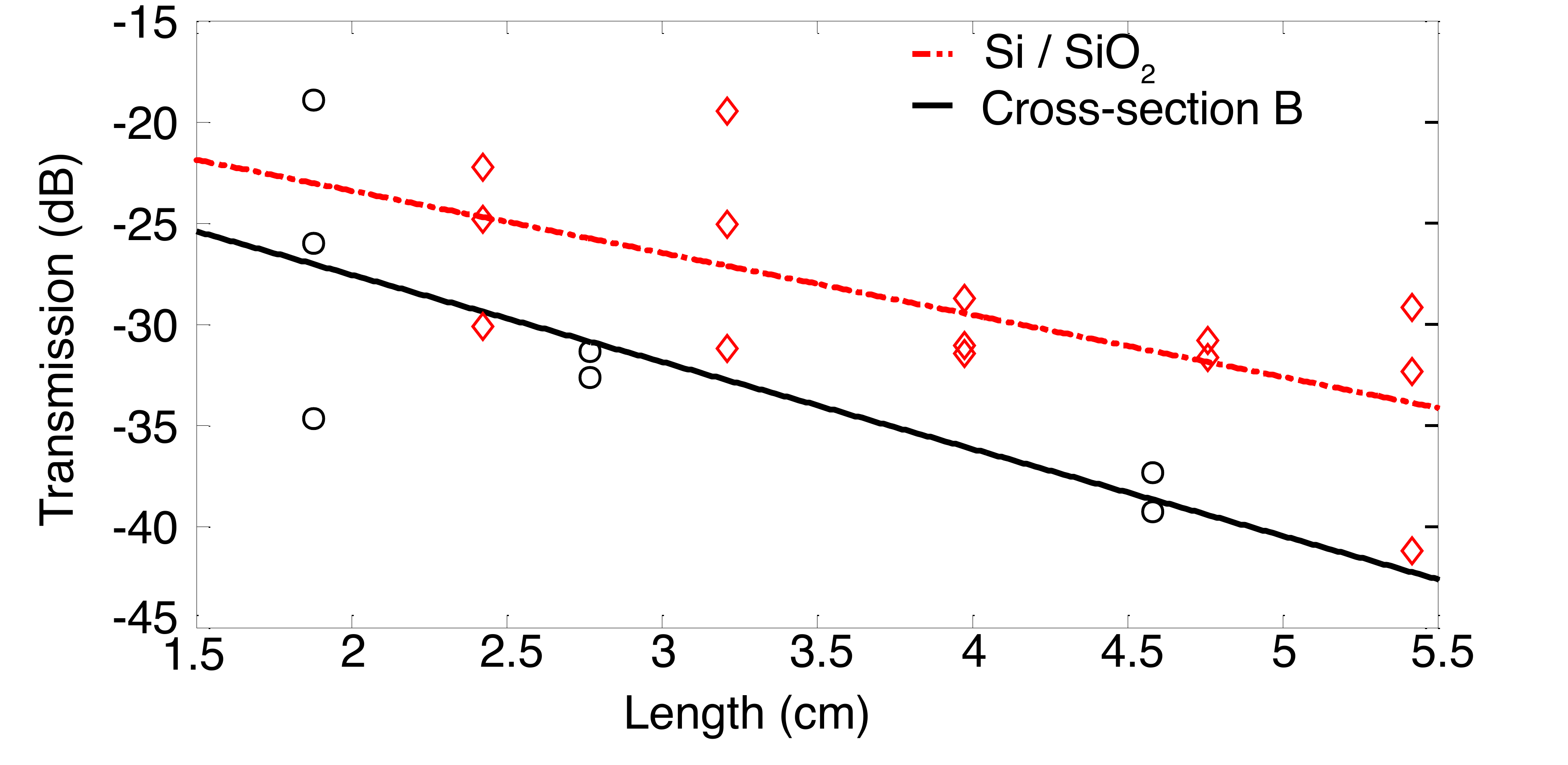}}
	\caption{Transmission versus length of waveguides and bends (paper-clip structures) at the wavelength of $1.55\ \mathrm{\mu m}$ for mode cross-section `B' in the straight sections and cross-section `A' in the semi-circular bends. The propagation losses of test structures consisting of 650~nm Si waveguides with $\text{SiO}_2$ cladding are also shown. The numerical fit is shown by the lines.}
	\label{F4}
\end{figure}

Photonic circuits can be assembled from the basic building blocks of waveguides and directional couplers. Hybrid waveguides of length between 1.8 cm and 4.6~cm were fabricated with the majority of the length consisting of waveguides using cross-section B. Cross-section A was used at the semi-circular bends, in order to fit the longer waveguides into a compact footprint using ``paperclip'' structures. The longest waveguide involves 10 transitions between cross-sections A and B. Since the edge facets were roughly diced and not polished, the fiber-to-chip insertion loss was high (about 9.5~dB without index-matching liquid) and non-uniform. As shown in Fig.~\ref{F4}, by measuring transmission versus length across several chips, we were able to build up an ensemble of measurements from which a propagation loss of $4.3 \pm 2.1$~dB/cm was extracted at 1550 nm wavelength, with minor changes across the wavelength range 1530~nm to 1570~nm. At this time, we are unable to separate the bending loss from the straight waveguide propagation loss, but the former is expected to be very small based on the large bending radius of $25\ \mathrm{\mu m}$. Numerical simulations reported in Fig.~\ref{F2}b suggest that the bending loss should drop to less than 0.005~dB per $90^\circ$ bend for bending radius greater than $3\ \mathrm{\mu m}$ using cross-section A. 

By way of comparison, the propagation loss of the same Si waveguides with $3\ \mathrm{\mu m}$ $\text{SiO}_2$ top-cladding (rather than LN) was measured to be $3.1\pm 2.1$ dB/cm, indicating that bonding and incorporating LN as the top cladding did not significantly worsen the propagation characteristics. In fact, the propagation loss is similar to that measured in rib Si photonic waveguides\cite{Vlasov:04} despite the thinner Si layer in our structures.

The loss values in these hybrid LN-Si waveguide using cross-section B are significantly lower than other reported values, e.g., 16~dB/cm in etched thin-film LN waveguides\cite{guarino2007electro} and 6 --10 dB/cm in thin-film LN waveguides (660~nm thickness, not too different from the 750~nm thickness used here) with oxidized titanium stripe\cite{li2015waveguides}. Although the loss is still an order-of-magnitude higher than that of traditional  in-diffused waveguides which have a much larger mode area, our circuits are also more than an order-of-magnitude more compact.  Furthermore, we expect that with improved Si waveguide fabrication (e.g., roughness reduction), and filling of the air pockets on the lateral sides of the Si rib with $\text{SiO}_2$ or a similar material, the overall propagation loss will decrease.  

\begin{figure}[t]
	\centerline{\includegraphics[width=0.5\linewidth]{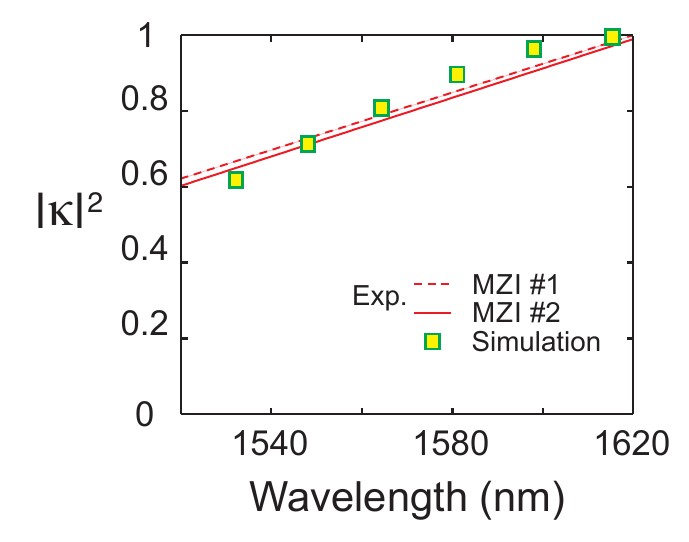}}
	\caption{Hybrid directional coupler characterization: Measurements (solid and dashed lines) and direct numerical simulation (yellow squares) of a hybrid LN-Si waveguide directional coupler.} 
	\label{F5}
\end{figure}

Directional couplers were defined by lithography in the Si layer, but act on the hybrid mode using cross-section A, with a gap of $0.4\ \mathrm{\mu m}$ between the waveguide edges. The typical length was only $150 \ \mathrm{\mu m}$, compared to a typical length of about 5~mm for directional couplers with diffused or ion-exchanged waveguides. Measurements are shown in Fig.~\ref{F5}, in good agreement with modeling using a numerical simulation (Lumerical software package) and the supermode theory which, unlike conventional coupled-mode theory, can be used for relatively high-index contrast waveguides\cite{6197696}. Because the mode in cross-section A resides primarily in Si, there was no measurable crosstalk at milliwatt power levels, unlike the traditional waveguide LN devices which are susceptible to photorefractive artifacts\cite{schmidt1980optically,Mueller:84}. However, long-term ($>$ 1 day) tests have not yet been performed since the chips are not fully packaged and we rely on manually-adjusted fiber coupling to the silicon waveguides.    

Using these building blocks, we designed and measured a hybrid LN-Si Mach-Zehnder interferometric structure consisting of two directional couplers, several bends and a path-length imbalance in one arm from which we were able to measure the effective modal indices of waveguides with cross-sections A and B. Also, on the same chip, outside the LN-bonded region, waveguides and devices may be designed as usual in silicon photonics. Because of the high refractive index of Si, light is mostly confined in Si for waveguides of cross-section `A' regardless of whether LN or  $\text{SiO}_2$ is the upper cladding. Therefore, waveguides with cross-section `A' and with $\text{SiO}_2$ replacing LN as the upper cladding (and with $\text{SiO}_2$ side- and lower-claddings) were used to define an MZI structure, similar to the hybrid MZI shown in Fig.~\ref{F1}. Although our test chip shown in Fig.~\ref{F1} was designed for LN covering nearly all of the Si surface, it is equally possible to  design ``mixed'' chips with smaller-sized LN pieces, which combine traditional Si photonic components with hybrid LN-Si photonic components on a monolithic platform. 

\section{Conclusion}
These results show that, even though LN is not a CMOS-compatible material, foundry-fabrication technologies can play a very useful role in a new generation of LN integrated optics, scaling to higher densities and increasing circuit complexities. While LN has always been a desirable material for integrated optics, it has not been possible in the past to make compact and complex waveguide circuits as is possible nowadays in Si photonics using precise and highly-repeatable DUV lithography. We have shown a suite of hybrid building blocks such as waveguides, couplers and interferometers from which more complex circuits can be built up. In order to optimize this platform, crystalline sub-micrometer thickness films of Si and LN must be intimately brought together, ideally with no intervening buffer layers. Here, we have shown chip-scale direct bonding of chips that are a few centimeters squared (the size of a typical field size of a DUV stepper lithography system), with enough bond strength to permit dicing and simple packaging for test and measurement. Similar techniques may also be applied to design optical circuits using other thin-film materials in place of LN, leveraging the advanced foundry fabrication capabilities of Si photonics as a waveguiding template for the hybrid modes, which eliminates the need to pattern the thin-films.   


\section*{Acknowledgment}
The authors acknowledge funding support from NSF ECCS 1307514 and the GOALI program. P.O.W. is grateful for support from the Department of Defense (DoD) through the National Defense Science \& Engineering Graduate Fellowship (NDSEG) Program. Sandia is a multiprogram laboratory operated by Sandia Corporation, a Lockheed Martin Company, for the United States Department of Energy's National Nuclear Security Administration under contract DE-AC04-94AL85000.

\cleardoublepage
\bibliographystyle{nature}
\bibliography{LiNbO3}

\cleardoublepage

\newpage

\newpage

\newpage

\newpage

\end{document}